\documentclass[aps,pre,twocolumn,amsfonts,floatfix,amssymb,superscriptaddress,showpacs]{revtex4}
\usepackage{placeins}
\usepackage{graphicx}
\usepackage{amssymb}
\usepackage{amsmath}
\usepackage{amsfonts}
\usepackage{hyperref}

\newcommand{\e}{\mathrm e}

\newcommand{\eq}[1]{(\ref{eq:#1})}


\date{\today}
\begin{document}

\title{
Universality of weak selection
}

\author{Bin Wu}
\email{bin.wu@evolbio.mpg.de}
\affiliation{Research Group Evolutionary Theory, Max-Planck-Institute for Evolutionary Biology, August-Thienemann-Str. 2, 24306 Pl{\"o}n, Germany}
\affiliation{Center for Systems and Control, State Key Laboratory for Turbulence and Complex Systems, College of Engineering, Peking University, Beijing 100871, China}
\author{Philipp M. Altrock}
\affiliation{Research Group Evolutionary Theory, Max-Planck-Institute for Evolutionary Biology, August-Thienemann-Str. 2, 24306 Pl{\"o}n, Germany}
\author{Long Wang}
\affiliation{Center for Systems and Control, State Key Laboratory for Turbulence and Complex Systems, College of Engineering, Peking University, Beijing 100871, China}
\author{Arne Traulsen}
\email{traulsen@evolbio.mpg.de}
\affiliation{Research Group Evolutionary Theory, Max-Planck-Institute for Evolutionary Biology, August-Thienemann-Str. 2, 24306 Pl{\"o}n, Germany}

\date{
\today}
\begin{abstract}
Weak selection, which means a phenotype is slightly advantageous over another,
is an important limiting case in evolutionary biology.
Recently it has been introduced into evolutionary game theory.
In evolutionary game dynamics, the probability to be imitated or to reproduce depends on the performance in a game.
The influence of the game on the stochastic dynamics in finite populations is governed by the intensity of selection.
In many models of  both unstructured and structured populations, a key assumption allowing analytical calculations is weak selection,
which means that all individuals perform approximately equally well.
In the weak selection limit many different microscopic evolutionary models have the same or similar properties.
How universal is weak selection for those microscopic evolutionary processes?
We answer this question by investigating the fixation probability and the average fixation time not only up to linear,
but also up to higher orders in selection intensity.
We find universal higher order expansions, which allow a rescaling of the selection intensity.
With this, we can identify specific models which violate (linear) weak selection results, such as the one--third rule of coordination games in finite but large populations.
\end{abstract}
\maketitle

\section{Introduction}

In evolutionary game theory the outcome of strategic situations determines the evolution of different traits in a population \cite{maynard-smith:1973to}.
Typically, individuals are hardwired to a set of strategies.
The performance in an evolutionary game determines the rate at which strategies spread by imitation or natural selection.
Due to differences in payoff, different strategies spread with different rates under natural selection.
In infinitely large well--mixed populations this is described by the deterministic replicator dynamics \cite{taylor:1978wv,hofbauer:1979mm,zeeman:1980ze,hofbauer:1998mm}.
In this set of non--linear
differential equations the intensity of selection,
which determines how
payoff affects fitness,
only changes the time scales, but not the direction of selection or the stability properties.
In finite populations fluctuations cannot be neglected
\cite{fogel:1998aa,ficici:2000aa,schreiber:2001aa,nowak:2004pw}.
The dynamics becomes stochastic: Selection drives the system into the same direction as the corresponding deterministic process, but sometimes the system can also move into
another direction.
The strength of selection determines the interplay between these two forces.
The absence of selective differences is called neutral selection:
Moving into one direction is as probable as moving into any other, independent of the payoffs.
If selection acts, the transition probabilities become
payoff dependent and thus asymmetric.
The asymmetry can be the same in each state (constant selection) or state dependent (frequency dependent selection).
In general, under frequency dependent selection the
probability that one
strategy replaces another
can be fairly complicated.
However, under the assumption of weak selection, some important insights can be obtained analytically
\cite{nowak:2004pw,traulsen:2006aa,ohtsuki:2006na,ohtsuki:2007pr,altrock:2009nj,kurokawa:2009aa,gokhale:2010pn,ohtsuki:2010aa}.
It has to be pointed out that these results do in general not carry over to stronger selection.

Weak selection describes situations in which the effects of payoff differences are small,
such that the evolutionary dynamics are mainly driven by random fluctuations.
This approach has a long standing history in population genetics \cite{kimura:1968aa,ohta:2002aa}.
In evolutionary biology, a phenotype is often found to be slightly advantageous over another phenotype \cite{akashi:1995ge,charlesworth:2007pn}.
Further, a recent experiment suggests that some aspects of weak selection are reflected in human strategy updating in behavioral games \cite{traulsen:2010pn}.
In the context of evolutionary game dynamics, however,
weak selection has only recently been introduced by Nowak et al.~\cite{nowak:2004pw}.
The definition of weak selection is unambiguous in the case of constant selection,
but there are different ways to introduce such a limit under frequency dependent selection \cite{traulsen:2010fy}.

In the simplest case, frequency dependence can be introduced by an evolutionary game between two types $A$ and $B$.
In a one shot interaction (where strategies are played simultaneously) a type $A$ interacting with another type $A$ receives payoff $a$, two interacting $B$ types get $d$ each.
Type $A$ interacting with $B$ gets $b$, whereas $B$ obtains $c$.
This symmetric $2\times2$ game can be described by the payoff matrix
\begin{align}\label{eq:Pmatrix}
\bordermatrix{
& A & B \cr
A & a & b \cr
B & c & d \cr}.
\end{align}
Let $i$ denote the number of $A$ individuals in a population of constant size $N$.
Under the assumption of a well-mixed population, excluding self--interactions,
the average payoffs for individuals of either type are given by
\begin{align}
\pi_A &= a \frac{i-1}{N-1} + b \frac{N-i}{N-1}\label{eq:MPayoffA},\\
\pi_B &= c \frac{i}{N-1} + d \frac{N-i-1}{N-1}\label{eq:MPayoffB},
\end{align}
These expectation values are the basis for
the evolutionary game.
In the continuous limit $N\to\infty$, the state of the system is characterized by the fraction of $A$ individuals $x=i/N$.
The dynamics are typically given by the replicator dynamics $\dot x = x (1-x) (\pi_A - \pi_B)$, which has the trivial equilibria $\hat x=0$ and $\hat x=1$.
Additionally, there can be a third equilibrium between zero and one, given by $x^\ast=(d-b)/(a-b-c+d)$.
In finite populations, the probabilistic description does not allow the existence of equilibrium points anymore.
Moreover, the invariance of the replicator dynamics under rescaling of the payoff matrix \cite{hofbauer:1998mm} is lost
in finite population models.
Typically, the average payoffs are mapped to the transition probabilities to move from state $i$ to other states,
only $i=0$ and $i=N$ are absorbing states.
When only two types compete and there is only one reproductive event at a time this defines a birth--death process.
The transition probabilities from $i$ to $i+1$, and from $i$ to $i-1$ are then denoted by $T_i^+$ and $T_i^-$, respectively.
They determine the probability of the process to be absorbed at a certain boundary, usually called fixation probability,
as well as the average time such an event takes, termed average fixation time.

An important result of evolutionary game dynamics in
finite populations under weak frequency dependent selection is the one--third rule.
It relates the fixation probability of a single type $A$ individual, $\phi_1$, to the position of the internal equilibrium $x^\ast$ in a coordination game, i.e.~when $a>c$ and $d>b$.
If selection is neutral, we have $\phi_1=1/N$.
If the internal equilibrium is less than one third, $x^\ast<1/3$, then $\phi_1>1/N$.
Originally, this weak selection result has been found for large populations in the frequency dependent Moran process \cite{nowak:2004pw}.
Subsequently, the one--third rule has been derived from several related birth--death processes \cite{traulsen:2005hp,traulsen:2006bb,ohtsuki:2007aa},
and also for the frequency dependent Wright-Fisher process \cite{imhof:2006aa,traulsen:2006ab},
which is still a Markov process, but no longer a birth-death process.
In a seminal paper, Lessard and Ladret have shown that the one--third-rule is valid for any process in the domain of Kingman's coalescence \cite{lessard:2007aa},
which captures a huge number of the stochastic processes typically considered in population genetics.
Essentially, this class of processes describes situations in which the reproductive success is not too different between  different types.
Thus, the generality of the one--third-rule under linear weak selection is well established.
Here we ask a slightly different
question:
To which order can two birth--death processes be considered as identical under weak selection?
Some authors have considered higher weak selection orders for specific processes \cite{ross:2007aa,bomze:2008lr,huang:2010aa}.
We investigate two classes of birth--death processes,
a general pairwise
imitation process motivated by social learning and a general Moran process based on reproductive fitness.
In this light, we also discuss cases which violate the one--third rule.

The manuscript is organized in the following way:
In Sec.\ \ref{sec:FixProb} we compute the weak selection expansion of the fixation probability in a general case of our two classes of birth--death processes.
In Sec.\ \ref{sec:FixTime}, we perform the same calculations for the significantly more complicated fixation times.
In Sec.\ \ref{sec:Disc} we discuss our analytical results and conclude.
Some detailed calculations can be found in the Appendix.

\section{Probabilities of fixation}\label{sec:FixProb}

A birth--death process is characterized by the transition probabilities from each state $i$ to its neighboring states,
$T_i^+$ and $T_i^-$.
We assume that this Markov chain is irreducible on the interior states and we exclude mutations or spontaneous switching from one type to another.
Thus, the process gets eventually absorbed at $i=0$ or $N$.
For any internal state, the probability to hit $i=N$ starting from $0<i<N$, $\phi_i$, fulfills the recursion equation
$\phi_{i}=(1-T_i^+-T_i^-)\phi_{i}+T_i^-\phi_{i-1}+T_i^+\phi_{i+1}$ \cite{goel:1974aa,ewens:2004qe,nowak:2006bo}.
This recursion can be solved explicitly, respecting the boundary conditions $\phi_0=0$ and $\phi_N=1$.
For a single $A$ individual in a populations of $B$, the probability to take over the population is \cite{goel:1974aa,ewens:2004qe,nowak:2006bo}
\begin{align}\label{eq:FixProb01}
\begin{split}
		\phi_1=\frac
		{
		1
		}
		{
		1+\sum_{k=1}^{N-1}\prod_{i=1}^{k}\frac{T_i^-}{T_i^+}
		}.
\end{split}
\end{align}
In any model of neutral selection, the transition probabilities of the Markov chain fulfill $T_i^-/T_i^+=1$, and hence the respective fixation probability of a single mutant
amounts to $1/N$.

In this section we focus on the weak selection approximation of Eq.~\eq{FixProb01}.
We do this for two different approaches to evolutionary game theory: imitation dynamics and selection dynamics.
In the former case, strategy spreading is based on pairwise comparison and imitation, in the latter it results from selection proportional to fitness and random removal.
The most prominent examples are the Fermi process and the Moran process, respectively.

\begin{figure}[t]
\begin{center}
\includegraphics[width=0.9\linewidth]{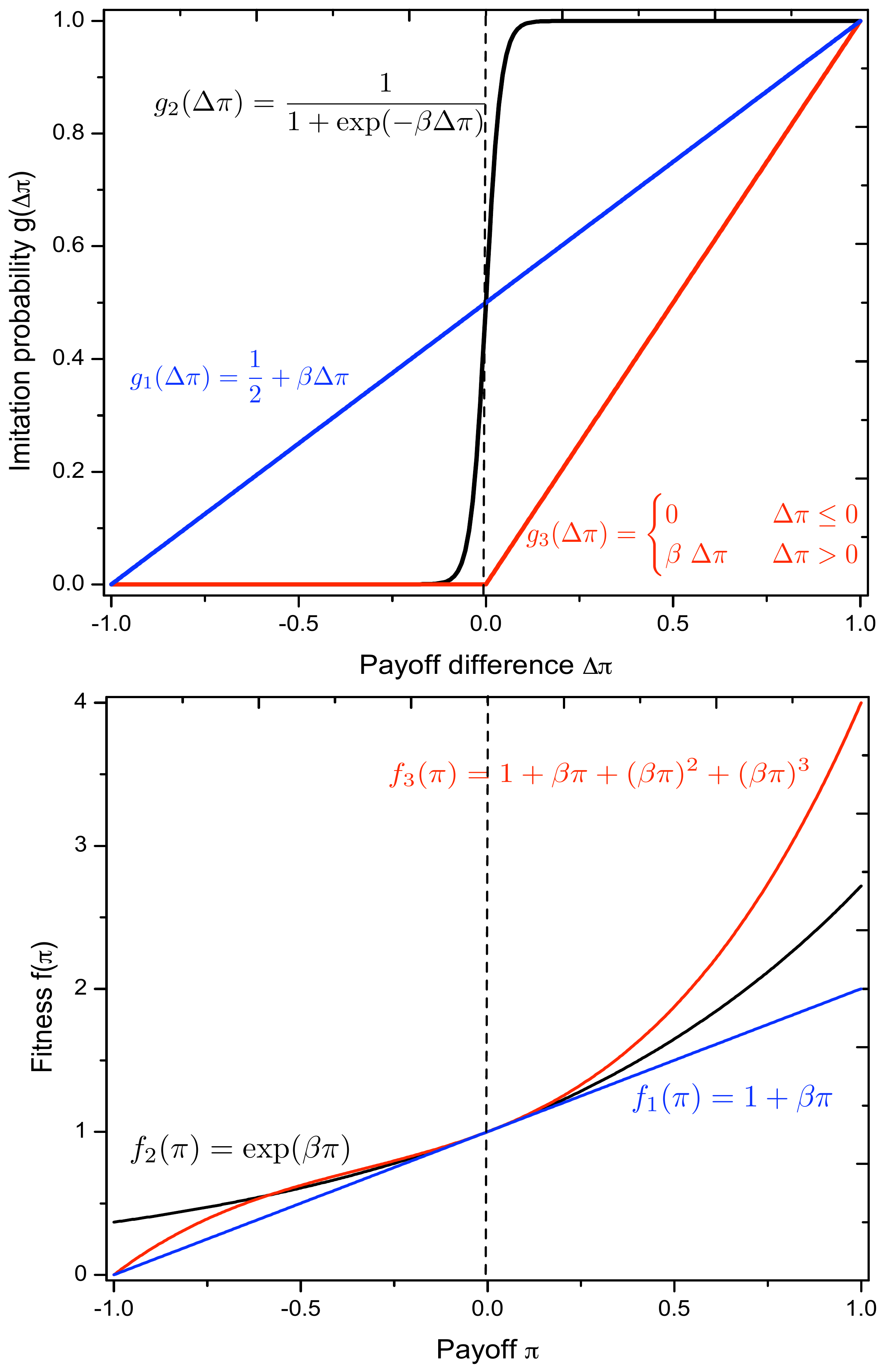}
\end{center}
\caption{
\label{fig1}
(color online)
Upper panel:
Pairwise comparison processes are characterized by the probability $g(\Delta \pi)$ to imitate the strategy of someone else based on the payoff difference $\Delta \pi$.
With increasing payoff difference, the imitation probability becomes higher, $g'(\Delta \pi) \geq 0$.
Weak selection implies a Taylor expansion at $\Delta \pi =0$.
Thus, it can only be invoked for functions that are differentiable in $0$.
The figure shows three examples of imitation probability functions,
$g_1(\Delta \pi)$ is a linear function
(selection intensity $\beta=0.5$)
and $g_2(\Delta \pi)$ is the Fermi function ($\beta=50$).
For the imitation function $g_3(\Delta \pi)$, a meaningful weak
selection limit does not exist since $g_3(\Delta \pi)$ is not differentiable in $0$.
Because $g_3(\Delta \pi)=0$ for $\Delta \pi <0$, the associated stochastic process would be stochastic in time, but deterministic in direction.
All through the manuscript, we focus on imitation functions that are differentiable
in $0$.
Lower panel:
Moran processes are characterized by a payoff to fitness mapping $f(\pi)$.
Fitness is a non--decreasing function of the payoff, $f'(\pi) \geq 0$.
The figure shows three examples for payoff to fitness mappings (selection intensity $\beta=1$ for all three functions).
}
\end{figure}

\subsection{Pairwise comparison}\label{ssec:PWC}

In a pairwise comparison process,
two individuals are chosen randomly to compare their payoffs from the evolutionary game, Eqs.~\eq{MPayoffA} and \eq{MPayoffB}.
One switches to the others strategy with a given probability, see Fig.\ \ref{fig1}.
If selection is neutral, this probability is constant.
If selection acts, the larger the payoff difference, the higher the probability that the worse imitates the better.
But typically there is also a small chance that the better imitates the worse.
Otherwise, only the strategy of the more successful individual is adopted.
This would lead to a dynamics that is stochastic in the time
spent in each interior state, but deterministic in direction \cite{traulsen:2006bb}.
Thus, given that all interior states are transient, the fixation probabilities are either 0 or 1 and
there is no basis to discuss a weak selection limit.

Selection is parameterized by the intensity of selection $\beta \geq 0$.
As a first example we consider the Fermi process \cite{blume:1993jf,szabo:1998wv,traulsen:2006bb}.
Let the two randomly selected individuals $X$ and $Y$ have payoffs $\pi_X$ and $\pi_Y$.
Then $X$ adopts $Y$'s strategy with probability
$g_{\text{Fermi}}(\pi_Y-\pi_X)=1/\left(1+\e^{-\beta (\pi_Y-\pi_X)}\right)$.
Thus, the transition probabilities of an evolutionary game with payoffs Eqs.~\eq{MPayoffA} and \eq{MPayoffB} are given by
\begin{align}\label{eq:Fermi01}
T_i^{\pm}=\frac{i}{N}\frac{N-i}{N}\frac{1}{1+\exp^{\mp\beta (\pi_A-\pi_B)}}.
\end{align}
The probability to stay in state $i$ is $1-T_i^--T_i^+$.
The Fermi process is closely related to Glauber dynamics \cite{glauber:1963aa}.
If we define individuals' energy as the exponential function of payoff, then the Fermi process can be mapped onto the Ising model.
The Fermi process has the comfortable property that the ratio of transition probabilities simplifies to $T_i^-/T_i^+=\e^{-\beta(\pi_A-\pi_B)}$,
such that the products
in Eq.~\eq{FixProb01} can be replaced by sums in the exponent.
Defining
$u=(a-b-c+d)/(N-1)$ and $v=(Nb-Nd-a+d)/(N-1)$, such that  $\pi_A-\pi_B=u\,i+v$, leads to
\begin{align}\label{eq:Fermi02}
\phi_1(\beta)
=\frac{1}
{1+\sum^{N-1}_{k=1}\exp\left\{-\beta\left[k^2\frac{u}{2}+k(\frac{u}{2}+v)\right]\right\}}.
\end{align}
For large $N$, the sum can be replaced by an integral, leading to a closed expression \cite{traulsen:2006bb}.
For weak selection, $N\beta\ll1$,
Eq. \eqref{eq:Fermi02} can be approximated by
\begin{align}\label{eq:Fermi03}
	\phi_1\approx\frac{1}{N}+\frac{(N-1)( (N+1)u+3v)}{6 N}\beta.
\end{align}
This can also be obtained directly from $T_i^-/T_i^+ \approx 1- \beta(\pi_A - \pi_B)$.
The fixation probability under weak selection is greater than in the neutral case if the term linear in $\beta$ is positive,
$Nu+u+3v>0$.
In particular, for a coordination game in a large population, this implies $x^\ast<1/3$.
Thus, natural  selection favors the mutant strategy, if the invasion barrier is less than one--third, which is the well known one--third rule \cite{nowak:2004pw,traulsen:2006bb,ohtsuki:2007aa,lessard:2007aa,bomze:2008lr}.
It holds when the fixation probability in a large but finite population can be approximated up to linear order in selection intensity.

\begin{figure}[t]
\begin{center}
\includegraphics[width=0.9\linewidth]{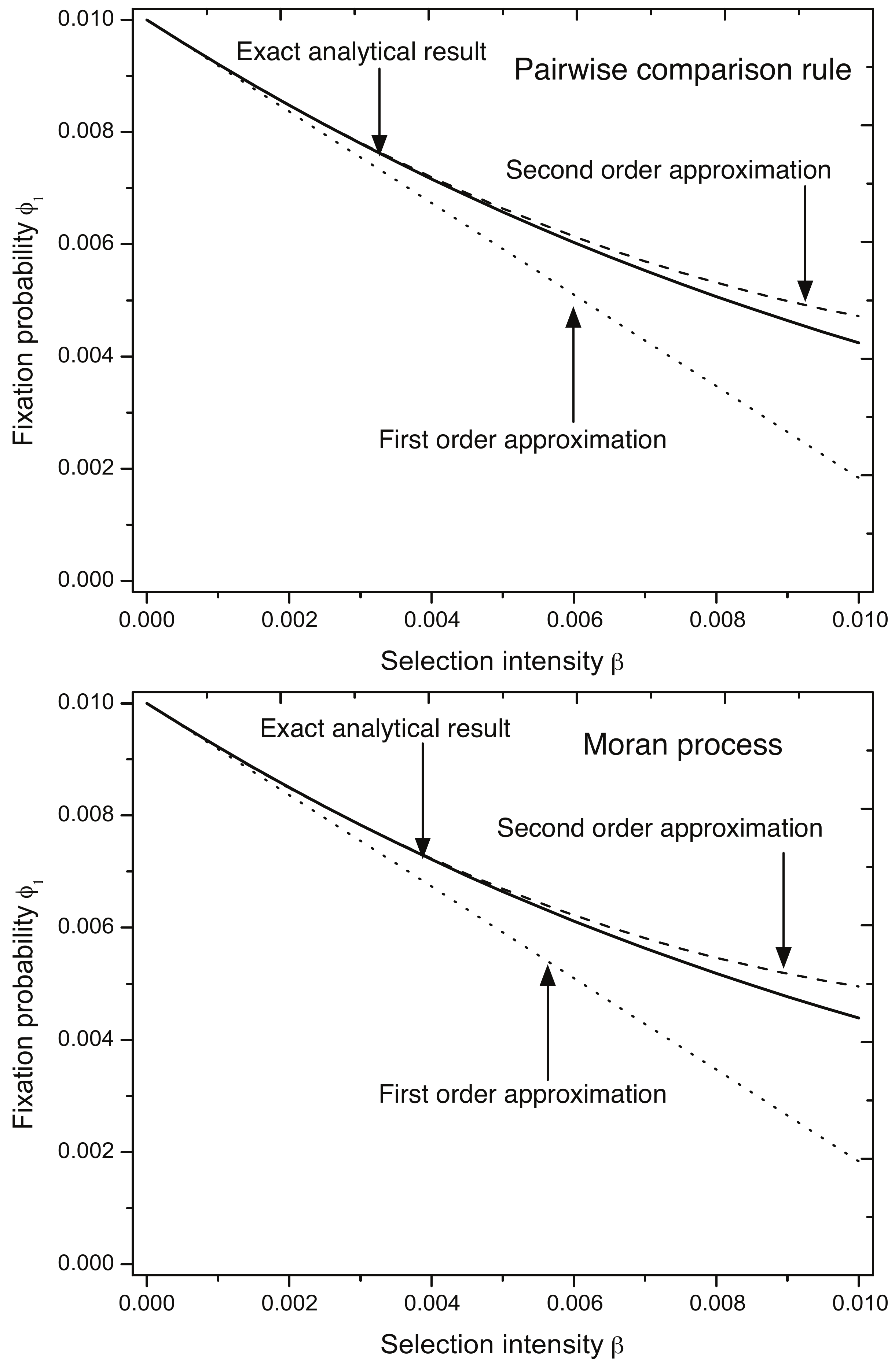}
\end{center}
\caption{
\label{fig2}
Approximation of the fixation probability of a single mutant under weak selection.
Upper panel:
Pairwise comparison process with the Fermi function $1/(1+\exp[-\beta \Delta \pi])$
as imitation function. As shown in the main text, up to second order the approximation is valid for any
imitation function $g(\beta \Delta \pi)$ after appropriate rescaling of the selection intensity $\beta$.
Lower panel:
Moran process with fitness as a linear function of the payoff, $f=1+ \beta \pi$.
Any other function leads to the same first order approximation after rescaling of $\beta$.
However, the second order depends on choice of the function transforming payoff to fitness.
Exact analytical results are numerical evaluations of Eq. \eqref{eq:FixProb01}.
(Parameters $N=100$, $\beta=1$,  $a=4$, $b=1$, $c=1$, and $d=5$ in both panels).
}
\end{figure}

Can we make general statements based on an expansion of $\phi_1$ concerning the probability of switching strategies,
$g(\Delta \pi)$?
In a general framework, the probability that $X$ switches to the strategy of $Y$, given the difference in their payoffs, $\Delta\pi=\pi_X-\pi_Y$, is governed by the intensity of selection.
We call $g(\Delta \pi)$ the imitation probability function of a general pairwise comparison process.
In a well mixed population, the transition probabilities read
\begin{align}\label{eq:Fermi04}
	T_i^{\pm}=\frac{i}{N}\frac{N-i}{N}g(\pm\beta\Delta\pi).
\end{align}
The larger the payoff difference, the more likely the worse individual switches to the strategy of the better.
Therefore the imitation function is nondecreasing, $g'(\Delta \pi) \geq 0$.
Additionally, if the payoffs of the two chosen individuals are equal, the neutral probability of switching is non-zero,
$g(0)>0$
(otherwise, the process does not allow a meaningful definition of weak selection
because it would always deterministically follow the direction of selection).
The fixation probability for this general pairwise comparison process can be expanded to the second order (see Appendix \ref{app:Aa})
\begin{align}\label{eq:Pairwise01}
	\phi_1\approx\frac{1}{N}+C_1\beta+C_2\beta^2,
\end{align}
where
\begin{align} \label{eq:Pairwise02}
\begin{split}
	C_1=\frac{(N-1)\left((N+1)u+3v\right)}{6N}
\frac{2g'(0)}{g(0)},
\end{split}
\end{align}
and
\begin{align} \label{eq:Pairwise03}
    C_2&=\left(
	u^2(N\!+\!1)(N\!+\!2)
	+15uv(N\!+\!1)
	+30v^2
	\right) \\
	& \times\frac{(N-1)(N-2)}{360}
	\left(\frac{2g'(0)}{g(0)}\right)^2.
	\nonumber
\end{align}
$C_1$ is proportional to the increase of the imitation function at $\Delta \pi=0$, see Fig.\ \ref{fig2}.
Note that for large $N$, $C_1>0$ is equivalent to $Nu+3v>0$, which for large $N$ further simplifies to
$x^{\ast}<1/3$.
Thus, the one--third rule holds for all pairwise comparison processes that fulfill $g'(0)\neq 0$, and $g(0)>0$.
Moreover, $C_1$ is proportional to $2g'(0)/g(0)$, while $C_2$ is proportional to the square of this quantity.
Thus, $2g'(0)/g(0)$ can be absorbed into the selection intensity by proper rescaling.
Therefore, the more rapid the increases of the imitation function at $\Delta \pi=0$,
the stronger is the sensitivity of the fixation probability to changes in average payoff.
For low switching probabilities in the neutral case, $\Delta \pi=0$,
we have a fixation probability that changes rapidly when the payoff difference becomes important, $\Delta \pi \neq0$.
While most previous models have either considered $g(0)=0$ (which lies out of the scope of our approach, because it does not lead to a reasonable definition of weak selection) or $g(0)=0.5$ (which is the default case), some authors have also
explored imitation functions with other values of $g(0)$.
For example, Szab{\'o} and Hauert have used  the imitation function $g(x)=1/(1+\e^{-x+\alpha})$, where $\alpha$ is a constant \cite{szabo:2002te}.
In this case $2g'(0)/g(0)=2/(1+\exp(-\alpha))$, thus, an increase in $\alpha$ is equivalent to an  increase in the (small) selection intensity.

Now it is straightforward to come up with an imitation function that leads to a violation of the one--third rule,
for example $g(\Delta \pi)=1/(1+\exp\{-\Delta \pi^3\})$.
Obviously, $g(\beta\Delta \pi)$ satisfies the conditions $g'(\beta\Delta \pi) \geq 0$, and $g(0)\neq0$.
Further, both the first and the second order expansions vanish.
Therefore, the fixation probability under weak selection can only be approximated as
\begin{align}\label{pairwise04}
	\phi_1\approx\frac{1}{N}+C_3\beta^3,
\end{align}
where $C_3$ can be derived in the same way as $C_1$ and $C_2$.
In special games, the sign of $C_3$ can also change at $x^\ast =1/3$, but in general
this will not be the case due to the complicated dependence of $C_3$ on $u$ and $v$.
In more general terms, the 1/3 rule is not sustained whenever the linear approximation of $g(\beta\Delta \pi)$ vanishes.

\subsection{Moran process}

In the frequency dependent Moran process
the payoff $\pi$, given by Eqs.~\eq{MPayoffA} and \eq{MPayoffB}, is mapped to fitness $f$, as illustrated in Fig.\ \ref{fig1}.
In each reproductive event, one individual is selected for reproduction (producing an identical offspring) proportional to fitness.
To keep the size of the population to the constant value $N$, a randomly chosen individual is removed from the population subsequently.
As in pairwise comparison processes,
the state $i$ can at most change by one per time step.

In the simplest case, fitness is a linear function of payoff.
With a background fitness of one, the fitnesses of type $A$ and $B$ read
$f_A=1+\beta\,\pi_A$, and $f_B=1+\beta\,\pi_B$, respectively.
The quantity $\beta\geq0$ serves as the intensity of selection.
Note that $\beta$ is bound such that fitness never becomes negative.
The probability that the number of $A$ individuals increases by one, $i\to i+1$, is given by
\begin{align}\label{eq:Moran01}
	T_i^+=\frac{i f_A}{if_A+(N-i)f_B}\frac{N-i}{N}.
\end{align}
The other possible transition, $i\to i-1$, occurs with probability
\begin{align}\label{eq:Moran02}
	T_i^-=\frac{(N-i)f_B}{if_A+(N-i)f_B}\frac{i}{N}.
\end{align}
When selection is neutral, $\beta=0$, we have $T_i^\pm=i(N-i)/N^2$.
Up to linear order in $\beta$ the Moran process has the same fixation probability as the Fermi process, Eq.~\eq{Fermi03},
such that in this approximation the one--third rule is fulfilled.
This is because under first order weak selection,  $T_i^-/T_i^+$ is again a linear function of the payoff difference.

In general, let fitness be {\it any} non-negative function of the product of payoff and selection intensity, $f(\beta\pi)$, which fulfills $f^\prime(\beta\pi)\geq0$.
For simplicity, we assume that the baseline fitness $f(0)$ is one.
The transition probabilities in a population with types $A$ and $B$
read
\begin{align}
	T_i^+&=\frac{i f(\beta\pi_A)}{i f(\beta \pi_A)+(N-i)f(\beta \pi_B)} \frac{N-i}{N} \label{eq:Moran03a},\\
	T_i^-&=\frac{(N-i)f(\beta\pi_B)}{i f(\beta \pi_A)+(N-i)f(\beta \pi_B)} \frac{i}{N}\label{eq:Moran03b}.
\end{align}
Note that $T_i^-/T_i^+ = f(\beta \pi_B) / f(\beta \pi_A)$.
Up to second order in $\beta$, the fixation probability of a single $A$ mutant in a population of $B$ is (see Appendix \ref{app:Ab})
\begin{align}\label{eq:Moran04}
\begin{split}
	\phi_1\approx\frac{1}{N}+D_1\,\beta+D_2\,\beta^2,
\end{split}
\end{align}
where
\begin{align}\label{eq:Moran05a}
D_1=(N-1)\frac{(N+1)u+3v}{6 N}\, f^\prime(0),
\end{align}
and
\begin{widetext}
\begin{align}\label{eq:Moran05b}
\begin{split}
	D_2=&
		\Bigg[u^2(N+1)(N+2) +15uv(N+1) +30v^2\Bigg]\frac{(N-1)(N-2)}{360}\, f'(0)^2\\
		&-\Bigg[(2a^2+4ab+4cd-10d^2)+(11d^2+2cd-c^2-3b^2-6ab-3a^2)N \\
		&+(a^2+2ab+3b^2-c^2-2cd-3d^2)N^2\Bigg] \frac{(N-1)}{24N^3}\left(f^\prime(0)^2-f^{\prime\prime}(0)\right),
\end{split}
\end{align}
\end{widetext}
with $u$ and $v$ as above.
Note that the first order term depends on payoff differences only, but the second order term also depends on the payoff values directly.
An example for such an approximation is shown in Fig. \ref{fig2}.
The first order term $D_1$ is proportional to the increase in fitness at $\pi=0$, $f'(0)$.
The first order term $D_1$ is proportional to $Nu+3v$ for large $N$.
Hence, the one--third rule holds for every Moran model for which $f'(0)$ does not vanish under weak selection.
Additionally, $f^\prime(0)$ can be absorbed into the selection intensity by rescaling:
Changing this rate is equivalent to changing the intensity of selection.
Note that this is not possible
with $D_2$, where not only the slope, but also the curvature of the fitness function at the origin plays a role.
However, when the exponential fitness function $f=\exp(\beta \pi)$ is employed \cite{traulsen:2008aa},
the second term of Eq.~\eq{Moran05b} vanishes.
This allows to incorporate $f'(0)$ into the selection intensity even for the second order term.

Again, we conclude the section with an example where the one--third is violated.
Consider the fitness function
$f(\beta \pi)=1+\beta^3 \pi^3$,
which obviously satisfies $f(0)=1$, and $f^\prime(\beta \pi)\geq0$.
Both, first and second order correction in $\beta$ vanish, $D_1=D_2=0$.
Therefore, the first non-trivial approximation of the fixation probability is
\begin{align}\label{eq:Moran06}
\begin{split}
	\phi_1\approx\frac{1}{N}+D_3\,\beta^3.
\end{split}
\end{align}
If $D_3$ changes sign at $x^\ast=1/3$, we recover the one--third rule.
This is only the case for very special games.
In analogy to the previous section, the general one--third rule does not hold anymore.

\section{Times of fixation}\label{sec:FixTime}

In this section we address the conditional fixation time $\tau_i^A$.
In a finite population of $N-i$ individuals of type $B$ and $i$ individuals of type $A$,
$\tau_i^A$ measures the expected number of imitation or birth--death events until the population consist of type $A$ only, under the condition that this event occurs.
In general, the probability $P_i^A(t)$ that after exactly $t$ events the process moved from any $i$ to $N$, which is the all $A$ state,
obeys the master equation
$P_i^A(t)\,=\,\left(1-T_i^+-T_i^-\right)P_i^A(t-1)+T_i^-P_{i-1}^A(t-1)+T_i^+P_{i+1}^A(t-1)$.
The average fixation time $\tau_i^A=\sum_{t=0}^{\infty}t\,P_i^A(t)/\phi_i$ is the stationary first moment of this probability distribution, resulting from a recursive solution of
$\phi_i\,\tau_i^A=\,\left(1-T_i^+-T_i^-\right)\phi_i\,\tau_i^A+T_i^-\phi_{i-1}(\tau_{i-1}^A+1)+T_i^+\phi_{i+1}(\tau_{i+1}^A+1)$.
In a similar way one can find $\tau_i^B=\sum_{t=0}^{\infty}t\,P_i^B(t)/(1-\phi_i)$, such that the total average lifetime of the Markov process amounts to $\phi_i\tau_i^A+(1-\phi_i)\tau_i^B$ \cite{goel:1974aa,karlin:1975xg,antal:2006aa}.
Following the previous section we restrict our analysis to the biologically most
relevant case $i=1$, which yields \cite{goel:1974aa,karlin:1975xg}
\begin{align}\label{eq:Time01}
	\tau_1^A = \sum\limits_{k=1}^{N-1}\,\sum\limits_{l=1}^{k}\,\frac{\phi_l}{T_l^+}\prod_{m=l+1}^{k}\frac{T_m^-}{T_m^+}.
\end{align}

Maruyama and Kimura \cite{maruyama:1974aa}, Antal and Scheuring \cite{antal:2006aa} as well as Taylor et al.\ \cite{taylor:2006jt} have shown
that the conditional fixation time of a single mutant of either type is the same, $\tau_1^A=\tau_{N-1}^B$.
This remarkable identity holds for any evolutionary birth--death process,
and is thus valid for any $2\times2$ game and for any selection intensity.
However, for $j>1$ we have $\tau_j^A\neq\tau_{N-j}^B$,
unless $\beta$ vanishes.
Since $\tau_1^A$ and $\tau_{N-1}^B$ are identical up to any order in $\beta$, we obtain
\begin{align}\label{eq:Time02}
	\left[\frac{\partial^n}{\partial\beta^n}\tau_{1}^A\right]_{\beta=0}=\left[\frac{\partial^n}{\partial\beta^n}\tau_{N-1}^B\right]_{\beta=0}
\end{align}
for any $n$.
This symmetry can help to obtain several properties of the expansion of the conditional fixation time, Eq.~\eq{Time01},
without brute force calculations.

\subsection{Pairwise comparison}\label{ssec:TimeP}

Let us first consider the fixation time in the special case of the Fermi process, Eq.~\eq{Fermi01}.
When the selection intensity vanishes, $\beta=0$, we have $\tau_1^A(0)=2N(N-1)$, \cite{altrock:2009nj,ewens:2004qe}.
When selection is weak, $N\,\beta\ll1$, the conditional fixation time is approximately
$\tau_1^A\approx
\tau_1^A(0)+\partial_\beta\tau_1^A(\beta)|_{\beta=0}\,\beta+\partial^2_\beta\tau_1^A(\beta)|_{\beta=0}\,\beta^2/2$.
For the Fermi process, the first order term is then given by \cite{altrock:2009nj}
\begin{align}\label{eq:TimeP01}
	\left[\frac{\partial}{\partial\beta}\tau_1^A\right]_{\beta=0}=-\,u\,N(N-1)\frac{N^2+N-6}{18},
\end{align}
where $u$ stems from $\pi_A-\pi_B=u\,i+v$, compare App.\ \ref{ssec:PWC}.
The first order expansion of $\tau_1^A$ is only proportional to the $i$ dependent term $u$ in this special case.
This can also be seen from a symmetry argument
\cite{taylor:2006jt,antal:2006aa}:
Since $\tau_1^A=\tau_{N-1}^B$,
the fixation time does not change under $a \leftrightarrow d$ and $b \leftrightarrow c$.
Since $u$, but not $v$, is invariant under this
exchange of strategy names,
$\tau_1^A$ can depend under linear weak selection only on $u$, but not on $v$.
The second order term of the conditional fixation time for the Fermi process yields
\begin{align}\label{eq:TimeP02}
	\left[\frac{d^2}{d\beta^2}\tau_1^A\right]_{\beta=0}=E_1\,u^2+E_2\,uv+E_3\,v^2,
\end{align}
where
\begin{align}
	E_1=&-\frac{(N\!-\!2)(N\!-\!1)N}{5400}(180-122N+177N^2+59N^3), \nonumber \\
	E_2=&-\frac{N^2(6-7N+N^3)}{18},\label{eq:TimeP03b}\\ \nonumber
	E_3=&\frac{1}{N}\,E_2.
\end{align}
Now, in contrast to the first order expansion Eq.~\eq{TimeP01}, both $u$ and $v$ enter.
An interesting relation is $E_3=E_2/N$.
In the following, we show that this is found for any pairwise comparison process and not only in the special case of the Fermi process.

For general pairwise comparison processes under neutral selection, the conditional fixation time is $\tau_1^A(0)=N(N-1)/g(0)$, where $g(0)>0$.
When selection acts,  Eq.~\eq{Fermi04}, the transition probabilities become dependent on the derivative of the imitation function,
$g^\prime(0) \geq 0$.
We are now interested in the imitation function's influence on the first and second order terms in $\beta$.
In general, the first order term in $\beta$ reads
\begin{align}
	&\frac{\partial}{\partial\beta}\tau_1^A
	=\!\sum_{|\alpha|=1}\sum_{k=1}^{N-1}\sum_{l=1}^{k}\,h_\alpha, \label{eq:TimeP04a}\\
      &h_\alpha=\left(\frac{\partial^{\alpha_1}}{\partial\beta^{\alpha_1}}\frac{1}{T_i^+}\right)
		\!\left(\frac{\partial^{\alpha_2}}{\partial\beta^{\alpha_2}}\phi_l\right)
		\!\left(\frac{\partial^{\alpha_3}}{\partial\beta^{\alpha_3}}\prod\limits_{m=l+1}^k\frac{T_m^-}{T_m^+}\right)\label{eq:TimeP04b}
\end{align}
with the multi--index $\alpha=(\alpha_1,\alpha_2,\alpha_3)$, $|\alpha |=\alpha_1+\alpha_2+\alpha_3$,
see App.\ \ref{app:Ba} for details of the calculation.
The general structure of this term is determined by $h_\alpha$, which is linear in $u$ and $v$, as $|\alpha|$ equals one.
Thus, $\partial_\beta\tau_1^A   |_{\beta=0}=F_1\,u+F_2\,v$ is also of this form, where $F_1$ and $F_2$ only depend on the population size $N$.
With the same symmetry argument as above, based on
\cite{taylor:2006jt,antal:2006aa},
we can conclude that $F_2=0$. This yields
\begin{align}\label{eq:TimeP05}
\begin{split}
	\tau_1^A=\tau_{N-1}^B\approx\frac{N(N-1)}{g(0)}+F_1\,u\,\beta.
\end{split}
\end{align}
We can now calculate the payoff independent term $F_1$ for any $g(\Delta \pi)$ from the special case $u=1$ and $v=0$,
which reads
\begin{align}\label{eq:TimeP06}
F_1=-\frac{g'(0)}{g(0)^2}\,N(N-1)\frac{N^2+N-6}{18}.
\end{align}
Here, $\beta$ can be rescaled by $g'(0)/g(0)^2$.
Changing $g'(0)$ or $g(0)$ is equivalent to changing the selection intensity appropriately.
In particular, when $u>0$,
which is true e.g.\ for coordination games such as the stag--hunt game \cite{skyrms:2003aa},
the conditional time it takes on average for a mutant type to take over decreases with the intensity of selection.
Moreover, for $a>c$ and $b>d$ in combination with $u<0$, a mutant which is always advantageous over the wild type
needs longer to reach fixation than a neutral mutant.
This phenomenon, termed stochastic slowdown in \cite{altrock:2010aa}, occurs in any imitation process, since Eq.\ \eqref{eq:TimeP05} only depends on $u$.

\begin{figure}[t]
\begin{center}
\includegraphics[width=0.9\linewidth]{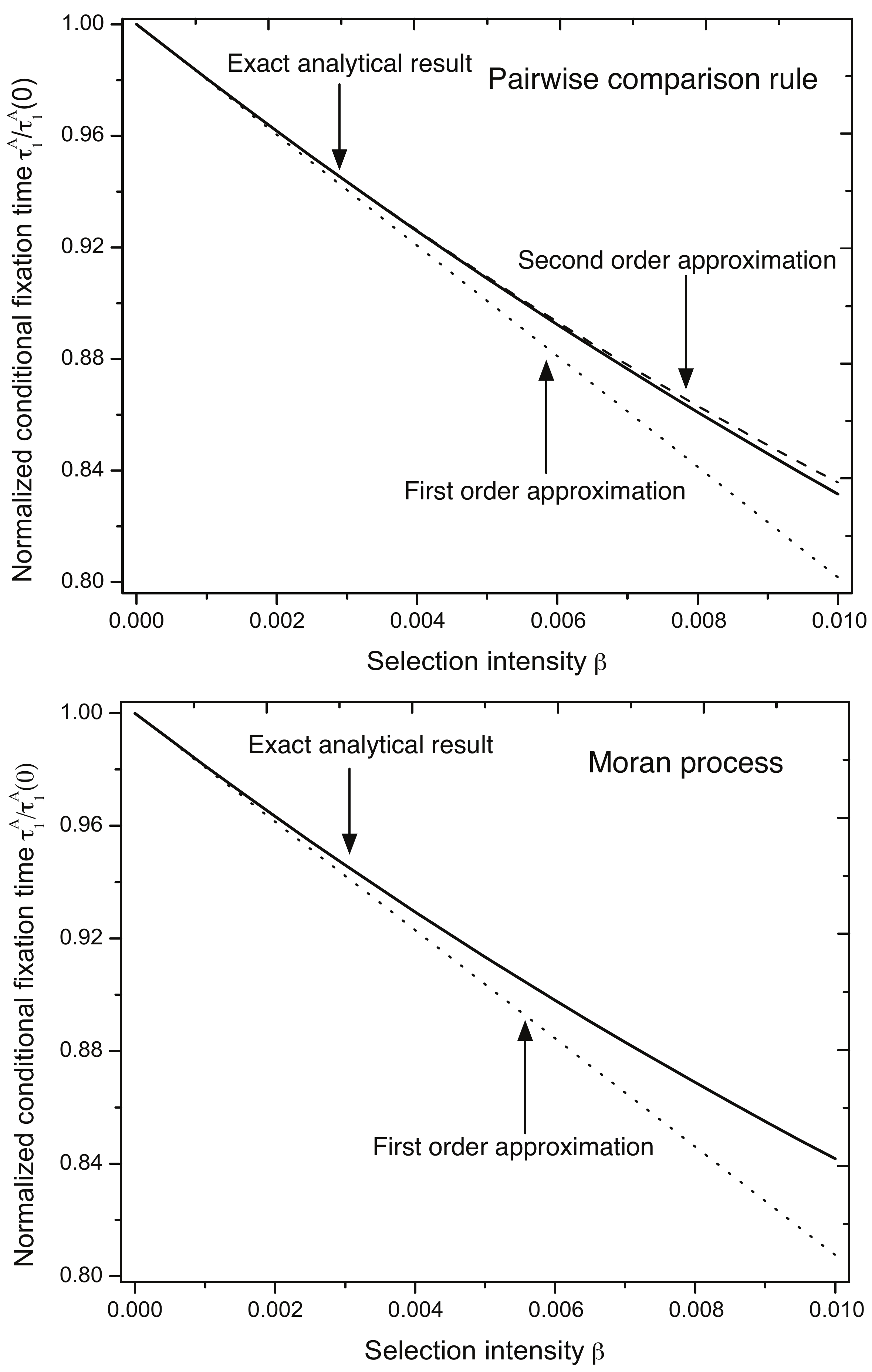}
\end{center}
\caption{
\label{fig3}
Weak selection approximation of the conditional fixation time of a single mutant,
the exact result is given by Eq.\ (\ref{eq:Time01}).
Upper panel:
The approximations are shown for the Fermi process, but they would be identical up to second order for any
other pairwise comparison process after appropriate rescaling of the selection intensity.
Lower panel:
For any Moran process  the first order approximation is independent of
the precise function mapping payoff to fitness (here it is linear).
Any higher order approximation depends on the details of the function.
Note that the first order approximation in the two panels is not identical due to a difference in the dependence
on population size $N$
(same parameters as in Fig.\ \ref{fig2})
}
\end{figure}

For the second order term in the expansion in $\beta$ we can write
\begin{align}\label{eq:TimeP07}
	\frac{\partial^2}{\partial\beta^2}\tau_1^A=\sum_{|\alpha|=2}\sum_{k=1}^{N-1}\sum_{l=1}^{k}\,h_\alpha,
\end{align}
$h_\alpha$ is of the form $G_1\,u^2+G_2\,uv+G_3\,v^2$.
Thus $\partial_\beta^2\tau_1^A|_{\beta=0}$ is also of this form, where the $G_i$'s only depend on $N$.
Again, we consider the transformation
$a \leftrightarrow d$ and $b \leftrightarrow c$
which corresponds to exchanging the names of the strategies.
For the transformed game, we obtain
$\partial_\beta^2\tau_{N-1}^B|_{\beta=0} = G_1\,u^2+G_2\,u\tilde v+G_3\,\tilde v^2$ with $\tilde{v}=(Nc-Na-d+a)/(N-1)$.
Using Eq.~\eq{Time02}, we obtain $G_2\,u(v-\tilde v)+G_3\,(v^2-\tilde v^2)=0$.
With $v+\tilde v=-N\,u$, we then get  $G_3=G_2/N$ --- the symmetry discussed above for a special case holds for any imitation function.
Eventually, the second order term in $\beta$ for general imitation probability is given by
\begin{align}\label{eq:TimeP08}
	\frac{\partial^2}{\partial\beta^2}\tau_1^A=G_1u^2+G_2uv+\frac{G_2}{N}v^2,
\end{align}
The special cases $u=1,\,v=0$, as well as $u=0,\,v=1$ allow to compute $G_1$ and $G_2$ explicitly.
Thus we have (see Appendix \ref{app:Ba})
\begin{widetext}
\begin{align}
\nonumber
	G_1=&-\frac{(N-2)(N-1)N}{5400}(180-122N+177N^2+59N^3)\left(\frac{2(g'(0))^2}{g(0)^3}\right)\\
	&-\frac{N^2(N-1)(2N-1)}{6} \left(\frac{g''(0)}{g(0)^2}\right)
	\label{eq:TimeP09a},\\
	G_2=&-\frac{N^2(6-7N+N^3)}{18}\left(\frac{2(g'(0))^2}{g(0)^3}\right)-N^2(N-1)\frac{g''(0)}{g(0)^2}\label{eq:TimeP09b}.
\end{align}
\end{widetext}
Obviously, Eq.(\ref{eq:TimeP08}) does not allow a rescaling of the intensity of selection.
Instead, the properties of the imitation function enter in a more intricate way.
An example of this approximation is shown in Fig. \ref{fig3}.

\subsection{Moran process}\label{ssec:TimeM}

To close this section, we consider the Moran process, where selection at birth is proportional to fitness and selection at death is random.
For neutral selection $\beta=0$, it is well known that  $\tau_1^A(0)=N(N-1)$ \cite{antal:2006aa,altrock:2009nj,ewens:2004qe}.
When selection is weak $\beta\ll1$, the conditional mean fixation time is approximately $\tau_1^A\approx\tau_1^A(0)+\partial_\beta\tau_1^A|_{\beta=0}\,\beta$.
For the Moran process with linear fitness function, $f_A=1+\beta\pi_A$, we have
$\partial_\beta\tau_1^A |_{\beta=0}=-u\,N^2(N^2-3N+2)/36$, compare \cite{taylor:2006jt,altrock:2009nj}.
The first order expansion of $\tau_1^A$ again depends only on $u$, but not on $v$.
This can be shown based on \cite{taylor:2006jt,antal:2006aa} or explicitly \cite{altrock:2009nj}.

With general fitness mapping $f(\beta\pi)$ with transition rates \eq{Moran03a} and \eq{Moran03b},
we have
\begin{align}\label{eq:TimeM01}
	\left[\frac{\partial}{\partial \beta}\tau_1^A(\beta)\right]_{\beta=0}=-f'(0)N^2\frac{N^2-3N+2}{36}u,
\end{align}
which allows a rescaling of the intensity of selection when $\tau_1^A$ is approximated up to linear order.

With general fitness function $f(x)$, it becomes unwieldy to calculate higher order terms in $\beta$.
However, the general calculations are similar to that of the general pairwise comparison rules.
Eq.~(\ref{eq:Moran05b}) reveals that already the second order expansion of the fixation probability $\phi_1$
with general fitness mapping is tedious in form.
Thus the equivalent terms for the fixation time $\tau_1^A$ are even more complicated and do not lead to further insight in this case.
Since it would be only an academic exercise to calculate them, we do not give them explicitly here.
It is clear that the weak selection approximation is not universal over a large class of processes in second order in the fixation times.

\section{Discussion}\label{sec:Disc}

In the past years, weak selection has become an important approximation in
evolutionary game theory \cite{nowak:2004pw,traulsen:2006aa,ohtsuki:2006na,ohtsuki:2007pr,altrock:2009nj,kurokawa:2009aa,gokhale:2010pn}.
Weak selection means that the game has only a small influence on evolutionary dynamics.
In evolutionary biology and population genetics, the idea that most mutations confer small
selective differences is widely accepted.
In social learning models, it refers to a case where imitation is mostly random, but there is a tendency to imitate others that are more successful.
Since weak selection is
 the basis of many recent results in evolutionary dynamics \cite{nowak:2006pw,ohtsuki:2006na,traulsen:2006aa,tarnita:2009df,antal:2009aa},
 it is  of interest how universal these results are.
It has been shown that they are remarkably robust and the choice of evolutionary dynamics has only a small impact in unstructured populations \cite{lessard:2007aa,antal:2009hc}.
In structured populations, however, the choice of evolutionary dynamics can have a crucial impact on the outcome
\cite{ohtsuki:2006na,ohtsuki:2006oz,tarnita:2009jx,tarnita:2009df,nowak:2010aa,roca:2009am,roca:2009aa}.
For example, for a prisoner's dilemma on a graph under weak selection, cooperation may be favored by a death birth process
while it is never favored by a birth death process.
In a well mixed population, however, the transition probabilities for those two processes are identical,
thus they lead to the same result.
However, in general spatial structure has a less pronounced effect under weak selection than under strong selection \cite{roca:2009am,roca:2009aa}.

We have addressed to what extent two evolutionary processes can be considered as identical by investigating the fixation probability and the fixation time.
For any given $2\times2$ payoff matrix,
we have considered two classes of evolutionary processes: Pairwise comparison and Moran processes.
An interesting special case is the Moran process with exponential fitness mapping, which is equivalent to the Fermi process (a special case of the pairwise comparison rule) in terms of fixation probabilities.

For the fixation probability, the first order term in the selection intensity always has the same form, given that it does not vanish.
In addition, regardless of the choice of imitation functions, two pairwise comparison processes are always identical up to second order
weak selection in the fixation probabilities. For Moran processes, an equivalent statement does not hold.
Recently, a paper has shown that in $3 \times 3$ games under weak selection,
the Fermi update rule can be quite different from the Moran process and the local update rule (an imitation process with linear imitation function \cite{traulsen:2005hp}),
while  the Moran process and the local update rule are more similar to each other \cite{bladon:2010ws}.
Our result shows that for weak selection in $2 \times 2$ games, these three processes can be mapped to each other by an appropriate rescaling
of the intensity of selection.

For the first order approximation of the average fixation time,
there are differences in the dependence on the system size, but all processes depend on the game in the same way.
This follows from a symmetry in fixation times \cite{taylor:2006jt,antal:2006aa}.
For higher orders in the intensity of selection,
a simple rescaling of the selection intensity does not exist for the fixation times and a general statement on the relation
between two processes cannot be made.

The robustness of weak selection results, i.e.\ the invariance to changes of the underlying stochastic process, found in the linear approximation is remarkable,
but follows from basic assumptions on evolutionary dynamics.
Moreover, the universality of weak selection breaks down
when higher order terms are discussed.

\section*{Acknowledgement}
B.W.\ gratefully acknowledges the financial support from China Scholarship Council (2009601286).
L.W.\ acknowledges support by the National Natural Science Foundation of China (10972002 and 60736022).
P.M.A.\ and A.T.\ acknowledge support by the Emmy-Noether program of the Deutsche Forschungsgemeinschaft.

\appendix
\begin{widetext}
\section{Third order expansion of the fixation probabilities}
\label{app:A}

Here, we expand the fixation probability $\phi_1$ for general birth--death processes up to the third order.
Let $\gamma_i=T_i^-/T_i^+$ and
\begin{align}
	\left[\frac{\partial^s}{\partial \beta^s}\gamma_i\right]_{\beta=0}=p_{si}.
	\end{align}	
Note that the first index of $p_{si}$ refers to the order of the derivative and the second index
gives the position in state space.
We expand Eq.(\ref{eq:FixProb01}) to the third order under weak selection $\gamma_i\approx1+p_{1i}\beta+p_{2i}\beta^2/2+p_{3i}\beta^3/6$.
Hence, we have
\begin{align}\label{GeneralFpTranProbExpansion}
	\prod_{i=1}^k\gamma_i\approx1
	&+\underbrace{\sum_{j=1}^kp_{1j}}_{L_{1k}}\beta
	+\underbrace{\left[\sum_{j=1}^k(p_{2j}-p_{1j}^2)+\left(\sum_{j=1}^kp_{1j}\right)^2\right]}_{L_{2k}}\frac{\beta^2}{2}\nonumber\\
	&+\underbrace{\left[\sum_{j=1}^kp_{3j}+3\left(\sum_{j=1}^kp_{1j}\right)\left(\sum_{s=1}^kp_{2s}\right)-3\sum_{j=1}^kp_{1j}p_{2j}\right]}_{L_{3k}}\frac{\beta^3}{6}.
\end{align}
Then the fixation probability can be written as
\begin{align}\label{GerneralFpMid}
	\phi_1&\approx
	\left(
	N
	+\beta\underbrace{\sum_{k=1}^{N-1}L_{1k}}_{Q_{1}}
	+\frac{\beta^2}{2}\underbrace{\sum_{k=1}^{N-1}L_{2k}}_{Q_{2}}
	+\frac{\beta^3}{6}\underbrace{\sum_{k=1}^{N-1}L_{3k}}_{Q_3}
	\right)^{-1}\\
\label{GeneralFpFinal}
& \approx \frac{1}{N}-\frac{Q_1}{N^2}\beta+\left[\frac{Q_1^2}{N^3}-\frac{Q_2}{2N^2}\right]\beta^2-\left[\frac{Q_1^3}{N^4}-\frac{Q_1Q_2}{N^3}+\frac{Q_3}{6N^2}\right]\beta^3.
\end{align}
This now serves as a starting point for our particular processes with certain choices of  $\gamma_i=T_i^-/T_i^+$ and
particular $p_{si}$ resulting from this.

\subsection{General pairwise comparison process}
\label{app:Aa}

For general switching probabilities in a pairwise comparison process,
we have
\begin{align}
p_{1i} &=-\frac{2 g'(0)}{g(0)}\Delta \pi_i, \\
p_{2i} &= \left(\frac{2g'(0)}{g(0)} \Delta \pi_i \right)^2 \\
p_{3i} &=-2\frac{6(g'(0))^3-3g(0)g'(0)g''(0)+g(0)^2 g'''(0)}{g(0)^3}  (\Delta \pi_i)^3.
\end{align}
Inserting these quantities into Eqs.(\ref{GeneralFpTranProbExpansion}) and (\ref{GerneralFpMid}) leads to
\begin{align}
Q_1&=-\frac{2g'(0)}{g(0)}\sum_{k=1}^{N-1}\sum_{i=1}^k\Delta \pi_i,\\
Q_2&=\left(\frac{2g'(0)}{g(0)}\right)^2\sum_{k=1}^{N-1}\left(\sum_{i=1}^k\Delta \pi_i\right)^2,\\
Q_3&=2\frac{6(g'(0))^3+3g(0)g'(0)g''(0)-g(0)^2g'''(0)}{g(0)^3}
 \sum_{k=1}^{N-1}\sum_{i=1}^k(\Delta \pi_i)^3 \nonumber\\
& -\frac{24(g'(0))^3}{g(0)^3}\sum_{k=1}^{N-1}\left(\sum_{i=1}^k\Delta \pi_i\right)\left(\sum_{s=1}^k(\Delta \pi_s)^2\right)\label{Q_3FroPair}.
\end{align}
$Q_1$ and $Q_2$ have been calculated in the main text.
Note that they only depend on $g'(0)/g(0)$, whereas $Q_3$ also depends on higher order derivatives of the imitation function.
Thus, two pairwise comparison processes that are identical in first order are also identical in second order.
Only in third order, differences start to emerge.

Let us briefly come back to our example of an imitation function that violates the $1/3$-rule,  $g(x)=(1+\exp(-x^3))^{-1}$.
In this case, we have $g(0)=1/2$, $g'(0)=g''(0)=0$ and $g'''(0)=3/2$.
Thus, both $Q_1$, and $Q_2$ vanish and the third order expansion of the fixation probability is
\begin{align}
	\phi_1\approx\frac{1}{N}+\frac{N-1}{60N}\left[(N+1)(3N^2-2)u^3+15(N+1)Nu^2v+30(N+1)uv^2+30v^3 \right]\beta^3.
\end{align}

\subsection{Moran processes}
\label{app:Ab}

For Moran processes with general fitness functions,
we have $p_{1i}=-f'(0)\Delta \pi_i$ and $p_{2i}=2(f'(0))^2\pi_A \Delta \pi_i -f''(0)(\pi_A+\pi_B)\Delta \pi_i$.
Inserting these quantities into Eqs.(\ref{GeneralFpTranProbExpansion}) and (\ref{GerneralFpMid}) leads to
\begin{align}
Q_1&=-f'(0)\sum_{k=1}^{N-1}\sum_{i=1}^k\Delta \pi_i\nonumber, \\
Q_2&=\left((f'(0))^2-f''(0)\right)\sum_{k=1}^{N-1}\sum_{i=1}^k(\pi_A^2-\pi_B^2)+(f'(0))^2\sum_{k=1}^{N-1}
\left(\sum_{i=1}^k\Delta \pi_i \right)^2.
\end{align}
Thus, the first and the second order expansion of the fixation probability of such processes are given by Eqs.(\ref{eq:Moran05a}) and (\ref{eq:Moran05b}), respectively.
In particular for $f(\pi)=1+\pi^3$,  both $p_{1i}$ and $p_{2i}$ vanish and  $p_{3i}=-6(\pi_A^3-\pi_B^3)$.
By Eq.(\ref{GerneralFpMid}), this yields
\begin{align}
\phi_1=\frac{1}{N}+\underbrace{\frac{1}{N^2}\sum_{k=1}^{N-1}\sum_{i=1}^k(\pi_A^3-\pi_B^3)}_{D_3}\beta^3+o(\beta^3)
\end{align}
where
$D_3
=\frac{1}{60N(N-1)^2}\Big(-3 c^2d (N-2)(1+N)(2N-1)-
 3cd^2(N-2)(N+1)(3N-4)+
 6a^2b(N-2)(N^2-2N+2)+
 a(a^2+3b^2)(N-2)(3N^2-6N+1)
 -c^3(1 + N)(3N^2-2)+
 2b^3(1+N-9N^2+6N^3)-
 d^3(N-2)(29 - 39 N + 12N^2)\Big)$.

\section{Times of fixation}
\label{app:B}

General expressions for the first and second order expansion of the fixation time for the birth--death process have been given in Eq(\ref{eq:TimeP04a}) and Eq.(\ref{eq:TimeP07}).
Based on these, we show the results for the general pairwise comparison rule first and then discuss the Moran process.

\subsection{General pairwise comparison process}
\label{app:Ba}

For the first order term of the fixation time, Eq.(\ref{eq:TimeP04a}), each $h_\alpha$ on the rhs.\ is proportional to $g'(0)/g^2(0)$.
Thus, the first order term of the fixation time is of the form $R g'(0)/g^2(0)$.
In particular, when $g(\Delta \pi)$ is the Fermi function, $g'(0)/g^2(0)$ is one.
Hence the first order of the fixation time for the Fermi process is $R$, cf.\ Eq.(\ref{eq:TimeP01}).
This leads to the first order expansion of the fixation time for general pairwise comparison rule, Eq.(\ref{eq:TimeP06}).

For the second order, we write Eq.~(\ref{eq:TimeP07}) explicitly as
\begin{align}
\frac{\partial^2}{\partial\beta^2}\tau_1^A&=\underbrace{\sum_{k=1}^{N-1}\sum_{l=1}^{k}\,h_{(2,0,0)}}_{K_1}
                                          +\underbrace{\sum_{k=1}^{N-1}\sum_{l=1}^{k}\,h_{(0,2,0)}}_{K_2}
              				+\underbrace{\sum_{k=1}^{N-1}\sum_{l=1}^{k}\,h_{(0,0,2)}}_{K_3}\nonumber \\
&                                          +\underbrace{2\sum_{k=1}^{N-1}\sum_{l=1}^{k}\,h_{(1,1,0)}}_{K_4}
                                         +\underbrace{2\sum_{k=1}^{N-1}\sum_{l=1}^{k}\,h_{(1,0,1)}}_{K_5}
                                          +\underbrace{2\sum_{k=1}^{N-1}\sum_{l=1}^{k}\,h_{(0,1,1)}}_{K_6}.
\end{align}
As shown in the main text, the second order term is of the form of $G_1u^2+G_2uv+\frac{G_2}{N}v^2$.
Letting $u=1$ and $v=0$ leads to
\begin{align}
\begin{split}
K_1&=\frac{N^2(N-1)(2N-1)}{6} \frac{2(g'(0))^2-g(0)g''(0)}{g(0)^3} \\
K_2&=-\frac{N^2(N-2)(N-1)(17+63N + 16 N^2)}{2700}\frac{2(g'(0))^2}{g(0)^3} \\
K_3&= \frac{N (-120 + 4 N + 350 N^2 - 65 N^3 - 290 N^4 + 121 N^5)}{1800}\frac{2(g'(0))^2}{g(0)^3} \\
K_4&=-\frac{N^3(N^2-1)}{12}\frac{2(g'(0))^2}{g(0)^3} \\
K_5&=\frac{N^3 (2 - 3 N + N^2)}{9}\frac{2(g'(0))^2}{g(0)^3}\\
K_6&=-\frac{ N^2 (2 + 25 N - 15 N^2 - 25 N^3 + 13 N^4)}{180} \frac{2(g'(0))^2}{g(0)^3}
\end{split}
\end{align}
after some tedious calculations using the identity $\sum_{k=1}^{M}\sum_{l=1}^k=\sum_{l=1}^{M}\sum_{k=l}^M$\cite{graham:1994aa}.
Summing these $K_i$'s leads to $G_1$ in Eq.(\ref{eq:TimeP09a}).
On the other hand, letting $u=0$ and $v=1$ yields
\begin{align}
\begin{split}
K_1&=N(N-1)\frac{2(g'(0))^2-g(0)g''(0)}{g(0)^3} \\
K_2&=\frac{N^2(N-1)(N-2)}{18}\frac{2(g'(0))^2}{g(0)^3} \\
K_3&=\frac{N(4N^3-15N^2+17N-6)}{18}\frac{2(g'(0))^2}{g(0)^3} \\
K_4&=-\frac{N^2(N-1)}{2}\frac{2(g'(0))^2}{g(0)^3} \\
K_5&= \frac{N(N-1)(N-2)}{2}\frac{2(g'(0))^2}{g(0)^3}\\
K_6&=-\frac{N^2(N-1)(N-2)}{3}\frac{2(g'(0))^2}{g(0)^3}
\end{split}
\end{align}
Adding these $K_i$'s  yields $G_2/N$ as in Eq.\ (\ref{eq:TimeP09b}).
Thus, the quantities in Eq.\ (\ref{eq:TimeP08}) are finally derived.

\subsection{Moran processes}
\label{app:Bb}

For Moran processes, the approach is fully equivalent to pairwise comparison processes.
However, the results do not only depend on payoff differences $u$ and $v$, but on the full payoff
matrix with entries $a$, $b$, $c$, and $d$.
This makes the calculations a matter of diligence and leads to
quite long expressions, but not to additional insights.
Thus, we do not give details of the derivation here.

\end{widetext}


\begin{thebibliography}{10}

\bibitem{maynard-smith:1973to}
J.~Maynard~Smith and G.~R. Price,
\newblock Nature {\bf 246}, 15 (1973).

\bibitem{taylor:1978wv}
P.~D. Taylor and L.~Jonker,
\newblock Math. Biosci. {\bf 40}, 145 (1978).

\bibitem{hofbauer:1979mm}
J.~Hofbauer, P.~Schuster, and K.~Sigmund,
\newblock J. Theor. Biol. {\bf 81}, 609 (1979).

\bibitem{zeeman:1980ze}
E.~C. Zeeman,
\newblock in {\it Lecture Notes in Mathematics} {\bf 819}, 471 (1980).

\bibitem{hofbauer:1998mm}
J.~Hofbauer and K.~Sigmund,
\newblock {\em Evolutionary Games and Population Dynamics} (Cambridge
  University Press, Cambridge, 1998).

\bibitem{fogel:1998aa}
G.~Fogel, P.~Andrews, and D.~Fogel,
\newblock Ecol. Model. {\bf 109}, 283 (1998).

\bibitem{ficici:2000aa}
S.~Ficici and J.~Pollack,
\newblock Effects of finite populations on evolutionary stable strategies.,
\newblock in {\em Proceedings GECCO}, edited by D.~Whitley {\em et~al.}, pp.
  927--934, Morgan-Kaufmann, San Francisco, 2000.

\bibitem{schreiber:2001aa}
S.~Schreiber,
\newblock Siam J. Appl. Math. {\bf 61}, 2148 (2001).

\bibitem{nowak:2004pw}
M.~A. Nowak, A.~Sasaki, C.~Taylor, and D.~Fudenberg,
\newblock Nature {\bf 428}, 646 (2004).

\bibitem{traulsen:2006aa}
A.~Traulsen and M.~A. Nowak,
\newblock Proc. Natl. Acad. Sci. USA {\bf 103}, 10952 (2006).

\bibitem{ohtsuki:2006na}
H.~Ohtsuki, C.~Hauert, E.~Lieberman, and M.~A. Nowak,
\newblock Nature {\bf 441}, 502 (2006).

\bibitem{ohtsuki:2007pr}
H.~Ohtsuki, M.~A. Nowak, and J.~M. Pacheco,
\newblock Phys. Rev. Lett. {\bf 98}, 108106 (2007).

\bibitem{altrock:2009nj}
P.~M. Altrock and A.~Traulsen,
\newblock New J. Physics {\bf 11}, 013012 (2009).

\bibitem{kurokawa:2009aa}
S.~Kurokawa and Y.~Ihara,
\newblock Proc. R. Soc. B {\bf 276}, 1379 (2009).

\bibitem{gokhale:2010pn}
C.~S. Gokhale and A.~Traulsen,
\newblock Proc. Natl. Acad. Sci. U.S.A. {\bf 107}, 5500 (2010).

\bibitem{ohtsuki:2010aa}
H.~Ohtsuki,
\newblock J. Theor. Biol. {\bf 264}, 136 (2010).

\bibitem{kimura:1968aa}
M.~Kimura,
\newblock Nature {\bf 217}, 624 (1968).

\bibitem{ohta:2002aa}
T.~Ohta,
\newblock Proc. Natl. Acad. Sci. USA {\bf 99}, 16134 (2002).

\bibitem{akashi:1995ge}
H.~Akashi,
\newblock Genetics {\bf 139}, 1067 (1995).

\bibitem{charlesworth:2007pn}
J.~Charlesworth and A.~Eyre-Walker,
\newblock Proc. Natl. Acad. Sci. U.S.A. {\bf 104}, 16992 (2007).

\bibitem{traulsen:2010pn}
A.~Traulsen, D.~Semmann, R.~D. Sommerfeld, H.-J. Krambeck, and M.~Milinski,
\newblock Proc. Natl. Acad. Sci. U.S.A. {\bf 107}, 2962 (2010).

\bibitem{traulsen:2010fy}
A.~Traulsen,
\newblock Evolution {\bf 64}, 316 (2010).

\bibitem{traulsen:2005hp}
A.~Traulsen, J.~C. Claussen, and C.~Hauert,
\newblock Phys. Rev. Lett. {\bf 95}, 238701 (2005).

\bibitem{traulsen:2006bb}
A.~Traulsen, M.~A. Nowak, and J.~M. Pacheco,
\newblock Phys. Rev. E {\bf 74}, 011909 (2006).

\bibitem{ohtsuki:2007aa}
H.~Ohtsuki, P.~Bordalo, and M.~A. Nowak,
\newblock J. Theor. Biol. {\bf 249}, 289 (2007).

\bibitem{imhof:2006aa}
L.~A. Imhof and M.~A. Nowak,
\newblock J. Math. Biol. {\bf 52}, 667 (2006).

\bibitem{traulsen:2006ab}
A.~Traulsen, J.~M. Pacheco, and L.~A. Imhof,
\newblock Phys. Rev. E {\bf 74}, 021905 (2006).

\bibitem{lessard:2007aa}
S.~Lessard and V.~Ladret,
\newblock J. Math. Biol. {\bf 54}, 721 (2007).

\bibitem{ross:2007aa}
A.~Ross-Gillespie, A.~Gardner, S.~A. West, and A.~S. Griffin,
\newblock Am. Nat. {\bf 170}, 331 (2007).

\bibitem{bomze:2008lr}
I.~Bomze and C.~Pawlowitsch,
\newblock J. Theor. Biol. {\bf 254}, 616 (2008).

\bibitem{huang:2010aa}
W.~Huang and A.~Traulsen,
\newblock J. Theor. Biol. {\bf 263}, 262 (2010).

\bibitem{goel:1974aa}
N.~Goel and N.~Richter-Dyn,
\newblock {\em Stochastic Models in Biology} (Academic Press, New York, 1974).

\bibitem{ewens:2004qe}
W.~J. Ewens,
\newblock {\em Mathematical Population Genetics} (Springer, NY, 2004).

\bibitem{nowak:2006bo}
M.~A. Nowak,
\newblock {\em Evolutionary Dynamics} (Harvard University Press, Cambridge, MA,
  2006).

\bibitem{blume:1993jf}
L.~E. Blume,
\newblock Games and Economic Behavior {\bf 5}, 387 (1993).

\bibitem{szabo:1998wv}
G.~Szab{\'{o}} and C.~T{\H{o}}ke,
\newblock Phys. Rev. E {\bf 58}, 69 (1998).

\bibitem{glauber:1963aa}
R.~J. Glauber,
\newblock J. Math. Phys. {\bf 4}, 294 (1963).

\bibitem{szabo:2002te}
G.~Szab{\'o} and C.~Hauert,
\newblock Phys. Rev. Lett. {\bf 89}, 118101 (2002).

\bibitem{traulsen:2008aa}
A.~Traulsen, N.~Shoresh, and M.~A. Nowak,
\newblock Bull. Math. Biol. {\bf 70}, 1410 (2008).

\bibitem{karlin:1975xg}
S.~Karlin and H.~M.~A. Taylor,
\newblock {\em A First Course in Stochastic Processes}, 2nd edition ed.
  (Academic, London, 1975).

\bibitem{antal:2006aa}
T.~Antal and I.~Scheuring,
\newblock Bull. Math. Biol. {\bf 68}, 1923 (2006).

\bibitem{maruyama:1974aa}
T.~Maruyama and M.~Kimura,
\newblock Evolution {\bf 28}, 161 (1974).

\bibitem{taylor:2006jt}
C.~Taylor, Y.~Iwasa, and M.~A. Nowak,
\newblock J. Theor. Biol. {\bf 243}, 245 (2006).

\bibitem{skyrms:2003aa}
B.~Skyrms,
\newblock {\em The Stag-Hunt Game and the Evolution of Social Structure}
  (Cambridge University Press, Cambridge, 2003).

\bibitem{altrock:2010aa}
P.~M. Altrock, C.~S. Gokhale, and A.~Traulsen,
\newblock Phys. Rev. E {\bf 82}, 011925 (2010).

\bibitem{nowak:2006pw}
M.~A. Nowak,
\newblock Science {\bf 314}, 1560 (2006).

\bibitem{tarnita:2009df}
C.~E. Tarnita, T.~Antal, H.~Ohtsuki, and M.~A. Nowak,
\newblock Proc. Natl. Acad. Sci. USA {\bf 106}, 8601 (2009).

\bibitem{antal:2009aa}
T.~Antal, H.~Ohtsuki, J.~Wakeley, P.~D. Taylor, and M.~A. Nowak,
\newblock Proc. Natl. Acad. Sci. USA {\bf 106}, 8597 (2009).

\bibitem{antal:2009hc}
T.~Antal, A.~Traulsen, H.~Ohtsuki, C.~E. Tarnita, and M.~A. Nowak,
\newblock J. Theor. Biol. {\bf 258}, 614 (2009).

\bibitem{ohtsuki:2006oz}
H.~Ohtsuki and M.~A. Nowak,
\newblock Proc. Roy. Soc. Lond. B {\bf 273}, 2249 (2006).

\bibitem{tarnita:2009jx}
C.~E. Tarnita, H.~Ohtsuki, T.~Antal, F.~Fu, and M.~A. Nowak,
\newblock J. Theor. Biol. {\bf 259}, 570 (2009).

\bibitem{nowak:2010aa}
M.~A. Nowak, C.~E. Tarnita, and T.~Antal,
\newblock Phil. Trans. Roy. Soc. London B {\bf 365}, 19 (2010).

\bibitem{roca:2009am}
C.~P. Roca, J.~A. Cuesta, and A.~S\'anchez,
\newblock Phys. Rev. E {\bf 80}, 046106 (2009).

\bibitem{roca:2009aa}
C.~P. Roca, J.~A. Cuesta, and A.~S\'anchez,
\newblock Physics of Life Reviews {\bf 6} (2009).

\bibitem{bladon:2010ws}
A.~J. Bladon, T.~Galla, and A.~J. McKane,
\newblock Phys. Rev. E {\bf 81}, 066122 (2010).

\bibitem{graham:1994aa}
R.~L. Graham, D.~E. Knuth, and O.~Patashnik,
\newblock {\em Concrete Mathematics}, second ed. (Addison-Wesley, 1994).

\end{thebibliography}
\end{document}